# Atomic Layer Deposition-Based Synthesis of Photoactive TiO$_2$ Nanoparticle Chains by Using Carbon Nanotubes as Sacrificial Templates

Shaoren Deng,[a,‡] Sammy W. Verbruggen,[b,c,‡] Zhanbing He,[d] Daire J. Cott,[e] Philippe M. Vereecken,[c,e] Johan A. Martens,[c] Sara Bals,[d] Silvia Lenaerts,*[b] and Christophe Detavernier*[a]

[a] *Department of Solid State Science, University Ghent, Krijgslaan 281/S1, B-9000 Ghent, Belgium. Fax: 00329 264 4996; Tel: 00329 264 4354; E-mail: christophe.detavernier@ugent.be*

[b] *Department of Bio-science Engineering, Research Group Sustainable Energy and Air Purification, University of Antwerp, Groenenborgerlaan 171, B-2020 Antwerp, Belgium. Fax: 00323 265 3225; Tel: 00323 265 3684; E-mail: silvia.lenaerts@ua.ac.be*

[c] *Center for Surface Chemistry and Catalysis, KU Leuven, Kasteelpark Arenberg 23, B-3001 Heverlee, Belgium*

[d] *Department of Physics, Electron Microscopy for Materials Science (EMAT), University of Antwerp, Groenenborgerlaan 171, B-2020 Antwerp, Belgium*

[e] *IMEC, Kapeldreef 75, B-3001 Leuven, Belgium*

## Abstract

Highly ordered and self supported anatase TiO$_2$ nanoparticle chains were fabricated by calcining conformally TiO$_2$ coated multi-walled carbon nanotubes (MWCNTs). During annealing, the thin tubular TiO$_2$ coating that was deposited onto the MWCNTs by atomic layer deposition (ALD) was transformed into chains of TiO$_2$ nanoparticles (~12 nm diameter) with an ultrahigh surface area (137 cm$^2$ per cm$^2$ of substrate), while at the same time the carbon from the MWCNTs was removed. Photocatalytic tests on the degradation of acetaldehyde proved that these forests of TiO$_2$ nanoparticle chains are highly photo active under UV light because of their well crystallized anatase phase.

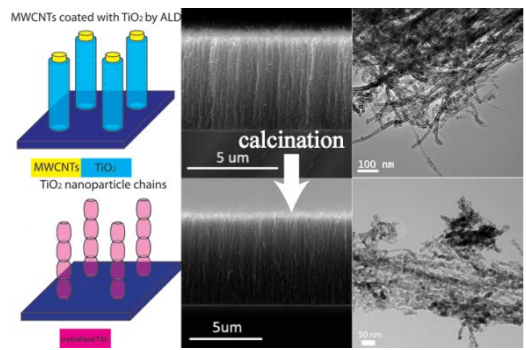

## INTRODUCTION

Due to its chemical stability, high photoactivity, suitable electronic properties and environmentally friendly nature, the semiconductor $TiO_2$ has attracted a lot of attention for various applications in modern technologies, e.g., in fuel cells, lithium ion batteries, gas sensors, solar cells, water splitting, and photocatalysis.[1-6] For most of these applications, nanostructured porous films with large accessible surface areas and optimized phase compositions and morphologies are required.[7,8] Many methods including electro-chemical anodization, template-assisted electrochemical deposition, sol-gel and hydro/solvothermal routes have been employed for the fabrication of $TiO_2$ nanostructures such as $TiO_2$ nanospheres, nano-tubes and nanowires.[9-11]

In this work, we present the controlled synthesis of one-dimensional $TiO_2$ nanoparticle chains anchored on a Si substrate. Our approach is based on the atomic layer deposition (ALD) technique. This thin film growth method is known to produce thin, conformal coatings with thickness control down to the atomic level, and has been proven successful for the synthesis and functionalization of nanostructured and highly porous materials.[12-16] Several authors have reported the ALD-based synthesis of $TiO_2$ nanotubes by using anodic alumina as a template.[17-19] Here, we use multi-walled carbon nanotubes (MWCNTs) as sacrificial templates. As illustrated in Scheme 1, a uniform and conformal $TiO_2$ coating is applied on a dense forest of MWCNTs. Calcination of the ALD-deposited amorphous $TiO_2$ into the photoactive anatase crystal structure results in burning away the underlying MWCNTs.[20,21] Simultaneously, a change in morphology of the $TiO_2$ layer occurs. Upon annealing, the dense $TiO_2$ coating transforms into interconnected anatase $TiO_2$ nanoparticles with a morphology of self supporting chains in the same orientation as the original MWCNTs. This morphology is drastically different from the cylindrical hollow tubes as typically obtained after chemical etching away of the anodic alumina templates. Finally, the combination of techniques used in our study results in a thin, porous and completely immobilized layer of $TiO_2$ nano particle chains that are perfectly suited as photocatalyst for applications in air pollution abatement.

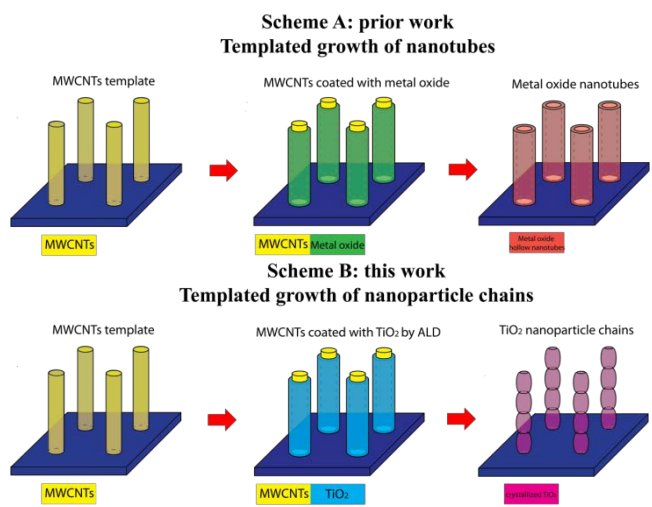

**Scheme 1**. Comparative overview of the synthesis route for obtaining metal oxide nanotubes (Scheme A) and TiO$_2$ nanoparticle chains (Scheme B) as discussed in this paper. In Scheme A, metal oxide were coated on MWCNTs by various method like sol-gel and electrospin. Metal oxide nanotubes were formed after the removal of the templates. In Scheme B, MWCNTs template is coated with TiO$_2$ by ALD. Subsequent annealing in air at 500°C results in the selective removal of the sacrificial CNT template and causes the transformation of the TiO$_2$ coating into vertical chains of nanoparticles.

## Experimental sections

### Growth of MWCNTs

The starting substrate was a Si wafer with a diameter of 200mm. MWCNTs were grown from a 1nm (nominal) Co catalyst layer. To avoid diffusion of the Co into the Si, a 70 nm TiN was first sputter deposited onto the Si surface (Endura PVD tool, Applied Materials, USA). MWCNTs were grown in a microwave (2.45 GHz) plasma enhanced chemical vapor deposition chamber (PECVD, TEL, Japan). In a typical experiment the Co catalyst layer was exposed to a NH$_3$ plasma for 5 mins to transform the film into active metal nanoparticles for CNT growth. Then a C$_2$H$_4$/H$_2$ mixture was flowed into the chamber at a temperature >550°C for 30mins. [22]

### TiO$_2$ ALD on MWCNTs and post deposition annealing

A 5 cm by 1.5 cm piece of grown MWCNTs was loaded into a homemade ALD tool with a base pressure in the low 10$^{-7}$ mbar range, as shown in Figure S1. [23] The sample was placed onto a heated chuck, and heated to 100°C. Tetrakis (dimethylamido) titanium (TDMAT) (99.999% Sigma-Aldrich) and O$_3$ generated by an ozone generator (Yanco Industries LTD) were alternately pulsed into the ALD chamber at pressures of 0.3 and 0.5mbar, respectively. The concentration of ozone in the flux was 145 µg/mL. The pulse (20 seconds) and pump time (40 seconds) were optimized to allow for a uniform coating of TiO$_2$ along the entire length of the MWCNTs and to prevent the occurrence of chemical vapor deposition type reactions. To remove the carbon, the $TiO_2$ coated MWCNTs sample was calcined in an oven at 500°C for 3 hours. The ramp rate was 1°C per minute and the ambient was air.

**Characterizations**

In situ and ex situ X-ray diffraction characterization (XRD) were measured in a home modified Bruker D8 system. (see Supporting Information) The morphology and elemental analysis of the cross-sections of the $TiO_2$ coated MWCNTs were carried out in a scanning electron microscope (FEI Quanta 200 F) equipped with energy dispersive x-ray analysis (EDX). The bright-field TEM images as well as the high-resolution transmission electron microscopy (HREM) images were obtained using a Tecnai G2 microscope operated at 200 kV. Both a Philips CM 20 and a Tecnai G2 microscope working at 200 kV were employed to get selected-area electron diffraction patterns (EDPs) for analysing the crystallographic phases of $TiO_2$ nanoparticles. For detailed description of the photocatalysis tests, please see Supporting Information

## Results and discussions

Uniform ALD on MWCNTs has been reported by depositing metal oxides directly onto the surface of MWCNTs.[24-27] The defective nature of MWCNTs being synthesized in a plasma environment was reported to enhance the nucleation of ALD processes.[28] In addition, it has been demonstrated that the $O_3$-based process for $Al_2O_3$ using trimethylaluminum (TMA) yields a better conformality on similar CNT arrays than the $TMA/H_2O$ ALD process.[29] Therefore, the choice was made to perform ALD of $TiO_2$ using TDMAT and $O_3$ as precursors. Figures 1(a) and (b) show optical and scanning electron microscopy (SEM) images of a MWCNTs sample after coating with 100 cycles of the $TiO_2$ ALD process. The transmission electron microscopy (TEM) image in Figure 1(c) illustrates that the ALD coating resulted in a $TiO_2$ layer surrounding most of the MWCNTs. A higher resolution TEM image in Figure 1(d) shows that the MWCNTs were uniformly coated by $TiO_2$. The original tube width was approximately 10 nm and broadened during ALD to ca. 16 nm, meaning that a 3 nm thick $TiO_2$ layer was deposited onto the MWCNTs, which corresponds to a growth rate of ca. 0.03 nm per ALD cycle. An Energy Dispersive X-ray spectroscopy (EDX) characterization was carried out during the SEM measurement to investigate the homogeneity of the $TiO_2$ coating throughout the entire MWCNTs layer. The inconspicuous variation of the Ti/C EDX signals ratio measured at distinct intervals along the length of the MWCNTs is indicative of a homogeneous distribution of $TiO_2$ in the entire layer (see Supporting Information). A supplementary X-ray fluorescence spectroscopy (XRF) test revealed that the amount of $TiO_2$ deposited by ALD was about 183 times larger than the amount of TiO$_2$ deposited on a planar Si substrate that experienced the same ALD process. Provided the growth per cycle on the flat reference sample and on the walls of the MWCNTs was similar, the MWCNTs offered an increase in surface area by a factor of about 183.[30] The very thin, uniform, perfectly controllable TiO$_2$ coating synthesized by ALD on the MWCNTs is strikingly different from other TiO$_2$ depositions on MWCNTs obtained via sol-gel, hydrothermal, solvothermal, electroplating, chemical vapor deposition and other chemical methods, where the coatings are often thicker than 3 nm, not conformal, coarser, uneven or less controllable.[31-38]

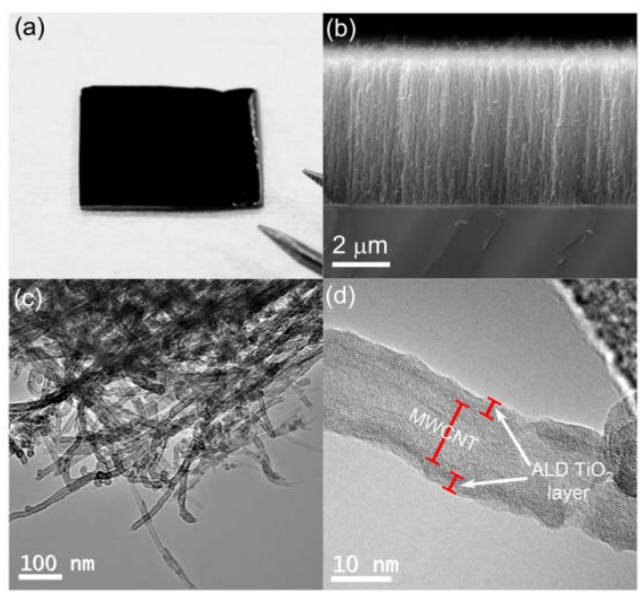

**Figure 1.** Optical image (a), SEM image (b) and TEM images (c) and (d) of 6 µm MWCNTs after 100 ALD cycles of TiO$_2$.

For applications such as photocatalysis and dye-sensitized solar cells, it is essential that the as-deposited amorphous TiO2 layer is transformed into a crystalline phase.[39, 40] The crystallization of a coated MWCNTs sample upon annealing was studied using in-situ x-ray diffraction (XRD). The sample was heated in air at a slow ramp rate of 1°C per minute from room temperature to 850°C, while irradiating it with a beam of Cu Kα radiation and continuously monitoring the diffraction peaks in the range of 15-35° (Figure 2 (a)). From these data it was concluded that crystallization of the TiO$_2$ layer on the MWCNTs into anatase phase initiates at about 400°C. Consequently, annealing of the wafer at 500°C for 3 h should result in a well developed TiO$_2$ anatase crystal structure that is suited for photocatalysis. The effect of such an annealing step is indicated by the XRD pattern depicted in Figure 2 (b). Note that the observed rutile phase is due to oxidation of the TiN layer beneath the MWCNTs, as further confirmed by a TEM study of the structure.[41] Elemental analysis of the annealed sample indicated that annealing in air resulted in the removal of the carbon from the sample, with only $TiO_2$ remaining, as illustrated in Figure 2 (c) and (d).

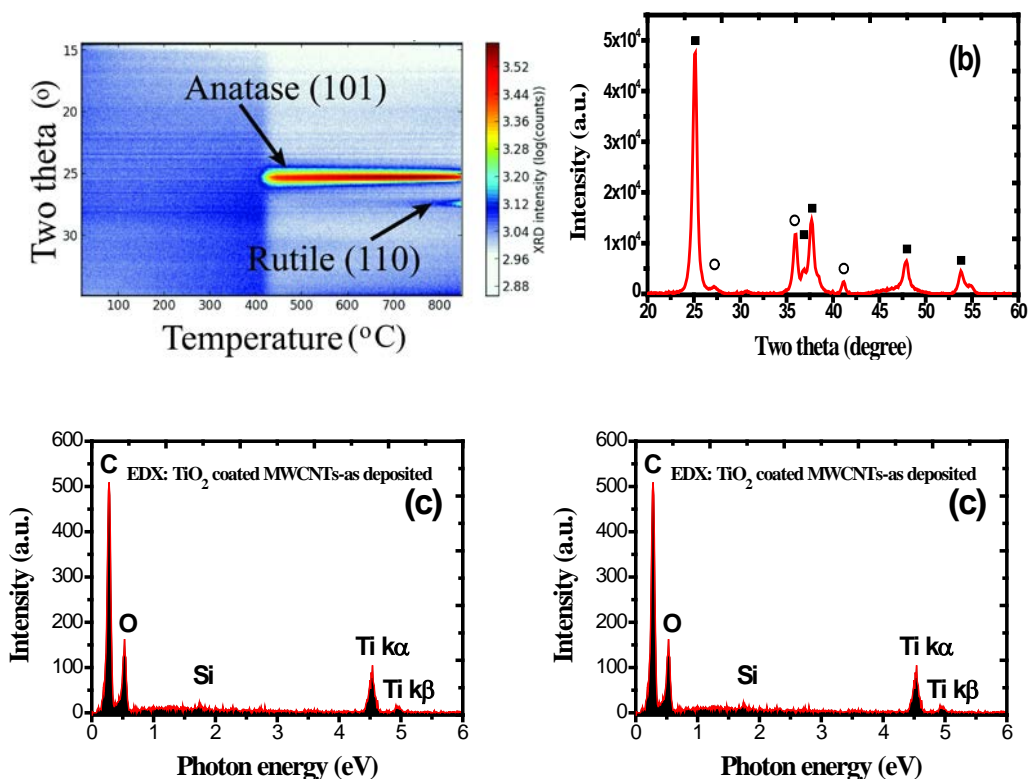

**Figure 2.** (a) In-situ XRD data of $TiO_2$ coated MWCNTs heated at 1°C·min-1 as a function of the temperature and diffraction angle $2\theta$. The appearance of anatase and rutile crystal phases is indicated. The colours in the diffraction pattern from blue to red represent the count of diffraction intensity from low to high. (b) XRD spectrum of $TiO_2$ coated MWCNTs after annealing in air at 500 °C for 3 hours. Anatase (■) and rutile (○) peaks are indicated. (c) and (d) show the EDX characterization of $TiO_2$ coated MWCNTs as-deposited (c) and after annealing in air at 500 °C for 3 hours (d).

Annealing in air of ultrathin coatings on flimsy supporting frameworks would normally cause a collapse of the structure.[42] However, due to the conformal nature of the $TiO_2$ ALD process on the MWCNTs forest, the $TiO_2$ coating transformed into a self-supporting nanostructure. Figure 3(a) shows an optical picture of the sample in Figure 1(a) after annealing in air at 500°C for 3 hours. Without apparent cracks or wrinkles at the macroscale, annealing clearly resulted in a color change from deep black to shiny purple blue. Given the black color of the MWCNTs, this color change can be related to the removal of carbon, as supported by the EDX results shown in Figure 2. The cross-sectional SEM picture in Figure 3(b) is very similar to Figure 1(b), indicating that the morphology at the microscale was substantially preserved and that calcination did not lead to a collapse of the structure, which is also a valid evidence for the excellent homogeneity of the ALD coating throughout the

MWCNTs layer. Further XRF characterization on this sample shows that this annealed porous $TiO_2$ film has a surface area increase by a factor of 137 compared to the planar substrate (i.e. 137cm$^2$ of $TiO_2$ surface per cm$^2$ of Si supporting substrate), showing that this porous film not only inherits the forest-like morphology of the MWCNTs template at the micro scale, but also retains as much as 75% of the surface area of the original MWCNTs forest.

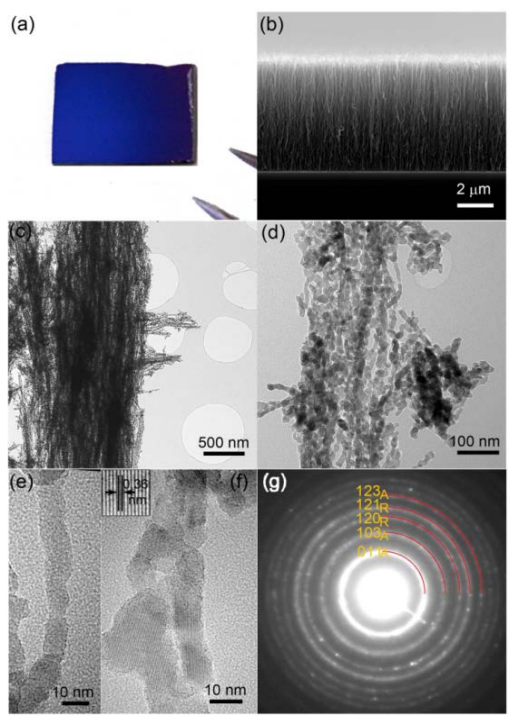

**Figure 3.** $TiO_2$ nanoparticle chains after annealing in air at 500 °C for 3 hours. (a) Optical image, (b) SEM image and (c-f) TEM images of the $TiO_2$ nanoparticle chains. The removal of the MWCNTs template and the transformation of the $TiO_2$ coating into nanoparticle chains was clearly revealed from both the low magnification bright-field TEM images in (c) and (d) and the HREM images in (e) and (f). The crystalline phase of the $TiO_2$ nanoparticles is demonstrated by the HREM images in (e) and (f) and the selected-area electron diffraction pattern in (g). The diffraction rings in (g) could be indexed both using the anatase and rutile structure, with a subscript A and R referring to anatase and rutile, respectively.

A more detailed investigation of the structure was performed using TEM. This study showed that calcination introduced a remarkable change in the morphology at the nano scale. During annealing the $TiO_2$ coating that originally surrounded the MWCNTs gradually transformed into a network of interconnected chains of nanoparticles (Figure 3(c) and (d)). A high resolution electron microscopy (HREM) image of such a particle revealed its crystalline nature (Figure 3(e)). A lattice spacing of 0.36 nm was distinguished, corresponding to the distance between anatase (011) planes (Figure 3(f)). The selected area electron diffraction pattern in Figure 3(g) indicates diffraction of (011)A, (103)A, (120)R, (121)R and (123)A crystal planes originating from both the anatase and rutile phase. A HREM study indicated that the $TiO_2$ nanoparticles only appear in the anatase phase, whereas the rutile phase could be related to oxidized TiN that was scratched from the surface during TEM sample preparation. The particles are fully crystalline throughout their entire cross-section and the mean $TiO_2$ particle size is ca. 12 nm (see supporting information). We suggest that the chain-like structure in the vertical direction could be formed because it originates from an ultrathin but continuous and conformal $TiO_2$ layer. When the continuity of the layer would be interrupted at several locations along the tubes, a substantial collapse of the structure is expected upon removal of the template. Furthermore, $TiO_2$ deposition at the numerous touching zones of the closely packed, flexible MWCNTs on the wafer is responsible for the lateral strengthening of the structure. It is worth noting that the formation of these metal oxide nanoparticle chains would likely not happen when either the metal oxide layer is too thick (>30nm) or the diameter of the MWCNTs is too large.[24, 43-44] Besides, by using ALD some groups reported that hollow metal oxides nanotubes or nanoribbons were formed when the organic templates like nanocellulose aerogels and peptides were removed by calcinations.[45-46] We postulate that during annealing in air, heat transmitted by the MWCNTs locally sintered the $TiO_2$ thin film and transformed the tubular coating into chains of nanoparticles after the MWCNTs were removed. In order to investigate the effect of annealing ambient, a control sample after the same ALD process was annealed in He ambient at 600°C for 3 hours. After annealing, as shown by the TEM picture in Fig. 4 (a), the $TiO_2$ thin coating on MWCNTs transformed into particles surrounding the MWCNTs, which remained intact during the annealing in inert ambient. However, due to the lack of oxygen and the carbonaceous template, only few $TiO_2$ nanoparticles were crystallized into the anatase phase while other $TiO_2$ nanoparticles preserved their amorphous nature.[47, 48] In Fig. 4 (b), XRD characterization on this sample further proved that the anatase [101] peak at 25.2° is much weaker compared to the sample annealed in air shown in Fig. 2 (b). The rutile peaks of $TiO_2$ apparent in Fig. 4 (b) are related to the annealing of the TiN layer that is present underneath the MWCNTs in He ambient at high temperature.[49]

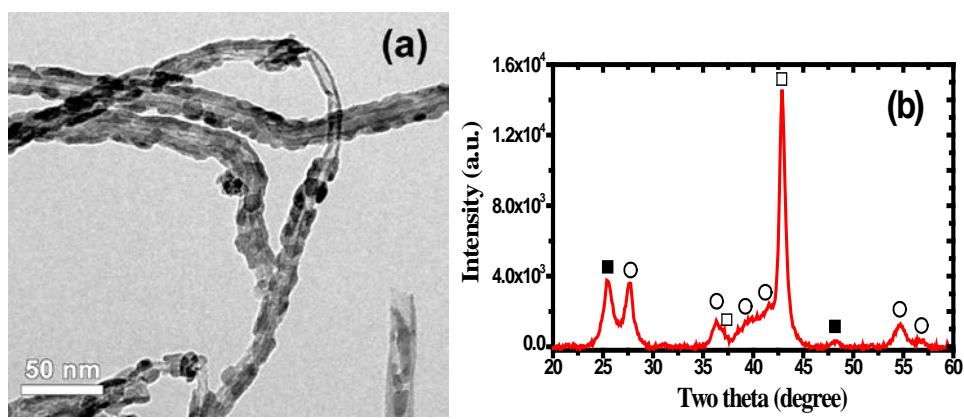

**Figure 4.** (a) TEM picture and (b) XRD spectrum of the control sample annealed in He ambient at 600 °C for 3 hours. Anatase (■), rutile (○) and TiN (□) peaks are indicated.

The obtained porous film, consisting of anchored anatase nano-particle chains, exhibits several merits for gas phase photocatalysis. The rigid structure provides a large surface area and a sufficiently porous arrangement to allow gas molecules to diffuse, adsorb and react on the active surfaces. Moreover the direct synthesis route of fixed TiO$_2$ nanoparticle chains offers an important benefit over common nanosized anatase powder particles that need additional immobilization efforts. Due to their nano size, nonfixed powder particles have the potential risk of jeopardizing human health.[50] As a control, a sonication treatment was applied to our film which proved that it is mechanically robust (see supporting information). These films could for instance have practical applications in the field of photocatalysis. As a proof of concept, a feasibility experiment involved testing a wafer of dimensions 5 cm by 1.5 cm towards photocatalytic acetaldehyde degradation in the gas phase. Acetaldehyde is an important source of indoor air pollution. Air spiked with 25, 52 and 80 ppmv of acetaldehyde was flowed over the film at a total flow rate of 400 cm$^3$·min$^{-1}$. After the adsorption-desorption equilibrium was established, the UV-A lamp was switched on. The evolution of the acetaldehyde concentration, as well as that of CO$_2$ as the degradation product was continuously monitored using on-line FTIR spectroscopy. More details on the photocatalytic test can be found in the Supporting Information. The change in the reactor outlet concentrations of acetaldehyde and CO$_2$ for an inlet concentration of 25 ppmv is displayed in Figure 5. From this it can be concluded that the moment the UV-A lamp is switched on, acetaldehyde is degraded into CO$_2$ (and water, not shown). The acetaldehyde degradation was determined to be 72%, 45% and 30% for the 25, 52 and 80 ppmv inlet concentrations, respectively. CO$_2$ is the main degradation product, which is evidenced by a carbon balance close to 100%. However, contrary to the TiO$_2$ nanoparticle chains with anatase phase, the control sample annealed in He showed no photo activity at all, which is due to the low proportion of crystalline nanoparticles in the film, as demonstrated by the TEM and XRD results discussed before.

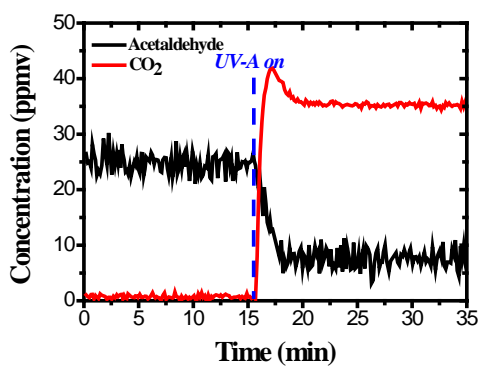

**Figure 5.** Reactor outlet concentrations of acetaldehyde (black curve) and CO$_2$ (red curve) during the photocatalytic test on TiO$_2$ nanoparticle chains. The moment the UV-A lamp is switched on is indicated with the dotted blue line.

## Conclusions

In conclusion, we introduced a novel ALD-based method to synthesize a forest of 1D anatase $TiO_2$ nanoparticle chains by using MWCNTs as sacrificial templates. The resulting self supported structure of fully crystallized nanoparticles offers a porous network with a large surface area, which is a desirable property for applications in e.g. photocatalysis, sensing and photovoltaics. Photocatalytic tests showed efficient degradation of acetaldehyde, a ubiquitous pollutant in indoor air, thus demonstrating that the self supported $TiO_2$ nanoparticle chains can be readily employed as an effective photocatalyst.

## Acknowledgements

The authors wish to thank the Research Foundation - Flanders (FWO) and UGENT-GOA-01G01513 for financial support. The authors acknowledge the European Research Council for funding under the European Union's Seventh Framework Programme (FP7/2007-2013)/ERC grant agreement n°239865-COCOON and N°246791-COUNTATOMS. JAM acknowledges the Flemish government for long-term structural funding (Methusalem).

## References

1. S. Y. Huang, P. Ganesan, and B. N. Popov, *ACS Catalysis.,* 2012, **2**, 825-831
2. S.W. Kim, T.H. Han, J. Kim, H. Gwon, H. S. Moon, S. W. Kang, S. O. Kim, and K. Kang, *ACS Nano*, 2009, **3**, 1085-1090
3. I. D. Kim, A. Rothschild, B. H. Lee, D. Y. Kim, S. M. Jo, and H. L. Tuller, *Nano. Lett.,* 2006, 6, 2009-2013
4. K. Zhu, N. R. Neale, A. Miedaner, and A. J. Frank, *Nano. Lett.,* 2007, **7**, 69-74
5. G. K. Mor, K. Shankar, M. Paulose, O. K. Varghese, and C. A. Grimes, *Nano. Lett.,* 2005, **5**, 191-195
6. A. Fujishima, X. Zhang, and D. A. Tryk, *Surface Science Reports,* 2008, **63**, 515-582.
7. S. Liu, H. Jia, L. Han, J. Wang, P. Gao, D. Xu, J. Yang, and S. Che, *Adv.Mater.,* 2012, 24, 3201-3204
8. J. L. Vivero-Escoto, Y. D. Chiang, K. C-W. Wu, and Y. Yamauchi, Sci. Technol. *Adv. Mater.,* 2012, **13**, 013003
9. P. Roy, S. Berger, and P. Schmuki, *Angew. Chem. Int. Ed.,* 2011, **50**, 2904-2939
10. H. E. Wang, H. Cheng, C. Liu, X. Chen, Q. Jiang, Z. Lu, Y. Y. Li, C. Y. Chung, W. Zhang, J. A. Zapien, L. Martinu, and I. Bello, *J. Power Source,* 2011, **196**, 6394-6399
11. X. Feng, K. Shankar, O. K. Varghese, M. Paulose, T. J. Latempa, and C. A. Grimes, *Nano Lett.,* 2008, **8**, 3781-3786
12. J. Dendooven, B. Goris, K. Devloo-Casier, E. Levrau, E. Biermans, M. R. Baklanov, K. F. Ludwig, P. Van Der Voort, S. Bals, and C. Detavernier, *Chem. Mater.,* 2012, **24**, 1992-1994
13. C. Bae, H. Shin, and K. Nielsch, *MRS Bulletin,* 2011, **36**, 887
14. M. Liu, X. Li, S. K. Karuturi, A. I. Y. Tok, and H. J. Fan, *Nanoscale,* 2012, **4**, 1522
15. C. Marichy, M. Bechelany, and N. Pinna, *Adv. Mater.,* 2012, **24**, 1017-1032
16. C. Detavernier, J. Dendooven, S. P. Sree, K. F. Ludwig, and J. A. Martens, *Chem. Soc. Rev.,* 2011, **40**, 5242
17. M. S. Sander, M. J. Cote, W. Gu, B. M. Kile, and C. P. Tripp, *Adv. Mater.,* 2004, **16**, 2052-2057.
18. M. Kemell, V. Pore, J. Tupala, M. Ritala, and M. Leskela, *Chem. Mater.,* 2007, **19**, 1816-1820
19. Y. C. Liang, C. C. Wang, C. C. Kei, Y. C. Hsueh, W. H. Cho, and T. P. Perng, *J. Phy. Chem. C,* 2011, **115**, 9498-9502
20. J. Sun, L. A. Gao, and Q. H. Zhang, *J. Mater. Sci. Lett.,* 2003, **22**, 339-341

21. A. B. Hungria, D. Eder, A. H. Windle, and P. A. Midgeley, *Catalysis Today,* 2009, **143**, 225-229
22. N. Chiodarelli, S. Masahito, Y. Kashiwagi, Y. Li, K. Arstila, O. Richard, D. J. Cott, M. Heyns, S. De Gendt, G. Groeseneken and P. M. Vereecken, *Nanotechnology,* 2011, **22**, 085302
23. Q. Xie, Y. L. Jiang, C. Detavernier, D. Deduytsche, R. L. Van Meirhaeghe, G. P. Ru, B. Z. Li, X. P. Qu, *J. Appl. Phys.,* 2007, **102**, 083521
24. Y. S. Min, E. J. Bae, K. S. Jeong, Y. J. Cho, J. H. Lee, W. B. Choi, and G. S. Park, *Adv. Mater.,* 2003, **15**, 1019-1022
25. C. F. Herrmann, F. H. Fabreguette, D. S. Finch, R. Geiss, and S. M. George, *App. Phys. Lett.,* 2005, **87**, 123110
26. D. S. Kim, S. M. Lee, R. Scholz, M. Knez, U. Gosele, J. Fallert, H. Kalt, and M. Zacharias, *App. Phys. Lett.,* 2008, **93**, 103108
27. S. Y. Liu, C. W. Tang, Y. H. Lin, H. F. Kuo, Y. C. Lai, M. Y. Tsai, H. Ouyang, and W. K. Hsu, *Appl. Phys. Lett.,* 2010, **96**, 231915
28. X. L. Li, C. Li, Y. Zhang, D. P. Chu, W. I. Milne and H. J. Fan, *Nanoscale Res. Lett.,* 2010, **5**, 1836-1840
29. N. Chiodarelli, A. Delabie, S. Masahito, Y. Kashiwagi, O. Richard, H. Bender. D. J. Cott, M. Heyns. S. De Gendt, G. Groeseneken, and P. M. Vereecken, *Mater. Res. Soc. Symp. Proc.,* 2011, 1283
30. J. Dendooven, S. Pulinthanathu Sree, K. De Keyser, D. Deduytsche, J. A. Martens, K. F. Ludwig, and C. Detavernier, *J. Phys. Chem. C,* 2011, **115**, 6605
31. W. G. Fan, L. Gao, J. Sun, *J. Am. Ceram. Soc.,* 2006, **89**, 731
32. R. Leary, A. Westwood, *Carbon,* 2011, **49**, 741
33. V. Krishna, S. Pumprueg, S. H. Lee, J. Zhao, W. Sigmund, B. Koopman, and B. M. Moudgil, *Process Saf. Environ. Prot.,* 2005, **83**, 393
34. S. Z. Kang, Z. Y. Cui, and J. Min, *Fullerenes Nanotubes and Carbon Nanostructures,* 2007, **15**, 81
35. Z. Z. Xu, Y. Z. Long, S. Z. Kang, and J. Mu, *J. Dispersion Sci. Technol.* 2008, **29**, 1150
36. G. M. An, W. H. Ma, Z. Y. Sun, Z. M. Liu, B. X. Han, S. D. Miao, Z. J. Miao, and K. L. Ding, *Carbon,* 2007, **45**, 1795
37. W. D. Wang, P. Serp, P. Kalck, and J. L. Faria, *J. Mol. Cata. A: Chem.,* 2005, **235**, 194
38. Y. Yao, G. Li, S. Ciston, R. M. Lueptow, and K. A. Gray, *Environ. Sci. Technol.,* 2008, **42**, 4952
39. Z. Ding, G. Q. Liu, and P. F. Greenfield, *J. Phy. Chem. B,* 2000, **104**, 4815-4820
40. T. S. Kang, A. P. Smith, B. E. Taylor, and M. F. Durstock, *Nano Lett.,* 2009, **9**, 601-606
41. B. Wang, K. Matsumaru, H. Ishiyama, J. Yang, K. Ishizaki, and J. Matsushita, *Advances in Technology of Materials and Materials Processing,* 2007, **13**, 7-13
42. H. J. Muhr, F. Krumeich, U. P. Schonholzer, F. Bieri, M. Niederberger, L. J. Gauckler, and R. Hesper, *Adv. Mater.,* 2000, **12**, 231-234
43. Y. Yang, L. Qu, L. Dai, T. S. Kang, and M. Durstock, *Adv. Mater.,* 2007, **19**, 1239
44. N. Du, H. Zhang, B. Chen, X. Ma, Z, Liu, J. Wu, and D. Yang, *Adv. Mater.,* 2007, **19**, 1641-1645
45. J. T. Korhonen, P. Hiekkataipale, J. Malm, M. Karppinen, O, Ikkala, and R. H. A. Ras, *ACS Nano,* 2011, **5**, 1967
46. S. W. Kim, T. H. Han, J. Kim, H. Gwon, H. S. Moon, S. W. Kang, S. O. Kim, and K. Kang, *ACS Nano,* 2009, **3**, 1085-1090
47. S. Deng, S. W. Verbruggen, S. Lenaerts, J. A. Martens, S. Van den Berghe, K. Devloo-Casier, W. Devulder, J. Dendooven, D. Deduytsche, and C. Detavernier, *J. Vac. Sci. Technol. A*, 2014, **32(1)**, 01A123
48. Q. Xie, J. Musschoot, D. Deduytsche, R. L. Van Meirhaeghe, C. Christophe, S. Van den Berghe, Y. L. Jiang, G. P. Ru, B. Z. Li, and X. P. Qu, *J. Electrochem. Soc.*, 2008, **155(9)**, H688-H692
49. F. H. Lu, and J. L. Lo, *J. Euro. Ceramic Soc.,* 2002, **22**, 1367-1374
50. L. Reijnders, *J. Hazardous Mater.,* 2008, **152**, 440